\documentclass[twocolumn,preprint,amsmath,amssymb, showpacs]{revtex4}
\usepackage{graphicx}
\usepackage{latexsym}
\usepackage{amssymb}
\usepackage{amsmath}
\newcommand{\YBCO}{ YBa$_{2}$Cu$_{3}$O$_{7}$ }
\newcommand {\LSMO} { La$_{2 / 3}$Sr$_{1 / 3}$MnO$_{3}$ }
\newcommand{\cuo}{CuO$_2$ }

\newcommand{\YPBCO}{Y$_{1-x}$Pr$_x$Ba$_2$Cu$_3$O$_7$}
\begin{document}

\title{Diverging Giant Magnetoresistance in the Limit of
Infinitely Conducting Spacer}

\author{Soumen Mandal$^1$}
\author{R. C. Budhani$^1$}
\email{rcb@iitk.ac.in}
\author{Jiaqing He$^2$}
\author{Y. Zhu$^2$}

\affiliation{$^1$Condensed Matter - Low Dimensional Systems laboratory, Department
of Physics, Indian Institute of Technology Kanpur,  Kanpur - 208016, India}
\affiliation{$^2$Materials Science Department, Brookhaven National Laboratory,
Upton, New York, 11973, USA}

\begin{abstract}
The relevance of pair-breaking by exchange and dipolar fields, and by injected spins
in a low carrier density cuprate Y$_{1-x}$Pr$_x$Ba$_2$Cu$_3$O$_7$ sandwiched between
two ferromagnetic \LSMO layers is examined. At low external field ($H_{ext}$), the
system shows a giant magnetoresistance(MR), which diverges deep in the
superconducting state. We establish a distinct dipolar contribution to MR near the
switching field(H$_c$) of the magnetic layers. At H$_{ext} \gg$ H$_c$, a large
positive MR, resulting primarily from the motion of  Josephson vortices and pair
breaking by the in-plane field, is seen.
\end{abstract}

\maketitle

Electron transport and magnetic ordering in ferromagnet (FM) - superconductor (SC)
heterostructures display a plethora of novel phenomena\cite{Gennes, Bergeret,
Keizer, Morgan} which acquire increasing richness in systems where the nature of the
FM and SC orders is exotic. Heterostructures of manganites and high temperature
superconducting cuprates offer such systems\cite{Melo}. The simplest structure which
potentially can display some of these phenomena is a trilayer where a SC film is
pressed between two ferromagnetic layers. Interestingly, such a system in the normal
state of the SC also constitutes the well-known spin valve in which two
ferromagnetic layers sandwich a non-magnetic(NM) metallic spacer\cite{book}.

The giant negative magnetoresistance (MR) seen in FM-NM-FM trilayers and multilayers
is related to asymmetric scattering of spin-up and spin-down electrons as they
cris-cross the spacer while diffusing along the plane of the
heterostructure\cite{book}. This flow of spin polarized charges is expected to
change profoundly when the spacer material becomes superconducting. Indeed, a large
MR has been seen by Pena et. al\cite{Visani} in La$_{0.7}$Ca$_{0.3}$MnO$_3$ - \YBCO
- La$_{0.7}$Ca$_{0.3}$MnO$_3$ trilayers in the narrow superconducting transition
region which they attribute to spin accumulation in YBCO when the FM layers are
coupled antiferromagnetically. These spins presumably cause depairing and hence a
large positive MR in accordance with the spin imbalance theory of Takahashi, Imamura
and Maekawa\cite{spinimbalance}.

In this letter we examine the relevance of pair breaking by dipolar and exchange
fields and injected spins in a low carrier density cuprate
Y$_{1-x}$Pr$_x$Ba$_2$Cu$_3$O$_7$ (YPBCO) which has insulating c-axis resistivity and
hence a poor spin transmittivity. We further address the issue of giant MR in \LSMO
- \YPBCO - \LSMO trilayers, in three distinctly illuminating ways which involve; i)
current density dependence of MR over a broad range of temperature below T$_c$, ii)
field dependence of MR when the magnetizations of \LSMO(LSMO) layers $\vec{M_1}$ and
$\vec{M_2}$ are parallel and fully saturated and, iii), dependence of MR on the
angle between current and field below and above the critical temperature (T$_c$).
These measurements permit disentanglement of the contributions of flux flow and pair
breaking effects in YPBCO, and the intrinsic anisotropic MR of LSMO layers to GMR in
FM-SC-FM trilayers, and establish a fundamental theorem which warrants diverging MR
in the limit of infinitely conducting spacer.

Epitaxial trilayers of LSMO-YPBCO-LSMO were deposited on (001) SrTiO$_3$. A
multitarget pulsed laser deposition technique was used to deposit the single layer
films and heterostructures as described in our earlier works \cite{Senapati}. The
thickness of each LSMO layer (d$_{LSMO}$) was kept constant at 30nm where as the
d$_{YPBCO}$ was varied from 30 to 100nm. The interfacial atomic structure of the
trilayers was examined with high resolution transmission electron microscopy (TEM)
at Brookhaven National Laboratory.

The high quality of plane LSMO films and of the films integrated in FM-SC-FM
heterostructures has been described in our previous reports\cite{Senapati}. We have
also investigated superconductivity in {\YPBCO} films as a function of Pr
concentration\cite{Li}.  As for single crystals\cite{Sandu}, the T$_c$ of the
films\cite{Li} also decreases with Pr concentration, and for $x \geq 0.55$, the
system has an insulating and antiferromagnetc (AF) ground state\cite{Tournier}. The
reduction in T$_c$ with x is presumed to be a consequence of lowering of the hole
concentration and their mobility due to the out-of-plane disorder caused by Pr ions.
Here we concentrate on x = 0.4 film because of its low carrier density and order
parameter phase stiffness\cite{Li}, both of which would enhance its susceptibility
to pair-breaking by spin polarized carriers injected from the LSMO.

Figure \ref{rtsem} shows the resistivity of a trilayer where the onset of hole
localization in YPBCO before superconductivity sets in is indicated by the rise in
resistivity near $\simeq$30 K. The inset of Fig.\ref{rtsem} shows a typical current
(I) - voltage (V) characteristic of the structure at 15K. The figure also shows a
cross-sectional TEM micrograph of the heterostructure. A sharp interface between
LSMO and YPBCO is seen in this atomic resolution image. While the manganite layers
are free of growth defects, we do see distinct stacking faults in the YPBCO, which
can be related to disorder due to difference in ionic radii of Y$^{3+}$ and
Pr$^{3+}$.

Figure \ref{rh15k} (Panel a) shows the magnetic field ($\vec{H}$) dependence of
magnetization ($\vec{M}$) at 10 K with $\vec{H}$ in the plane of the heterostructure
and parallel to the easy axis [110] of LSMO. Starting from a fully saturated
magnetic state at H$\simeq$ 180Oe, the M reaches a plateau over a field range of -90
to -130 Oe on field reversal. This is indicative of AF alignment of the $\vec{M}$
vectors of the top and bottom LSMO films. We have demonstrated earlier that the poor
c-axis conductivity of YBCO actually quenched the oscillatory part of the interlayer
exchange interaction and only an exponentially decaying AF-exchange remains in the
LSMO-YBCO-LSMO system\cite{Senapati}. The full cancelation of the moment
(M$\simeq$0) seen in the plateau also suggests that the two layers have equal
saturation magnetization (M$_s$). In the remaining three Panels of Fig. \ref{rh15k}
we show the in-plane resistance of the trilayer as a function of $\vec{H}$ coplanar
with the measuring current(I). Two values of the angle ($\theta$) between the I and
H have been chosen; in one case  $\theta = 0^o$ (Fig. \ref{rh15k}b) and for the
other two panels (c \& d), $\theta = 90^o$ but the magnitude of I is different.
While these measurements have been performed at several currents, only a few
representative field scans of MR are shown in Fig. \ref{rh15k}. The MR for both
$\theta = 90^o$ and $\theta  = 0^o$ configurations has two distinct regimes of
behavior. Starting from a fully magnetized state at 500 Oe and $\theta = 90^o$, the
MR first drops to a minimum as the field is brought to zero following a dependence
of the type $\sim \alpha$H + $\beta$H$^2$, where $\alpha = -6.8 \times 10^{-6}$ and
$\beta = 7.7 \times 10^{-8}$ for I = 0.8mA and $\theta = 90^o$. The MR shows a
step-like jump at the reversed field of $\sim40$ Oe where the magnetization switch
to AF configuration and remains high till $\vec{M_1}$ and $\vec{M_2}$ become
parallel again. One reversing the field towards positive cycle, a mirror image of
the curve is seen in the positive field quadrant. A remarkable feature of the MR is
its dependence on current I. The peak MR at 500 Oe drops from $\sim$80\% to 17\% on
increasing the current by a factor of two. The height of the MR curves remains
nearly the same when the magnetic field is rotated from  $\theta = 90$ to $\theta =
0$ with some differences in its shape. The pertinent factors which affect the MR of
such structures are; i) the MR in the normal state of YPBCO, ii) the explicit role
of superconductivity which is suppressed by the dipolar and exchange fields of the
FM layers, and by the spin polarized electrons injected from the FM layers, and
iii), a parasitic non-zero tilt of the sample away from parallel configuration which
will result in a high concentration of vortices in the superconducting spacer even
at low fields. These factors are addressed with the help of Fig. \ref{rhvart} where
we have plotted M(H) loop at 40 K (normal state). The AF alignment of $\vec{M_1}$
and $\vec{M_2}$ in the vicinity of zero-field persists in the normal state as well.
Figure \ref{rhvart} also shows the field-dependence of MR for $\theta = 90^o$ at a
few temperatures as the sample is taken from superconducting to normal state. A
striking drop in MR on approaching the normal state is evident in addition to a
noticeable change in its field dependence. At 40 K, it drops monotonically on
reducing the field from full saturation till the reverse switching field is reached
where it shows a small but discernible step-like increase followed by an
unremarkable field dependence in the negative field side. For I$\|$H ($\theta =
0^o$) the R(H) curve (not shown) reflects the anisotropic magnetoresistance (AMR) of
LSMO films.

It becomes clear from Fig. \ref{rh15k} and \ref{rhvart} that the H-dependence of MR
in these FM-SC-FM trilayers can be divided into two regimes, one covering the range
-150Oe $<$ H $<$ 150Oe where the reorientation of $\vec{M_1}$ and $\vec{M_2}$ is the
deciding factor, and at the higher fields where $\vec{M_1} \| \vec{M_2}$ and it goes
as $\sim \alpha$H + $\beta$H$^2$. While the MR in these regimes is intimately linked
with superconductivity of YPBCO, its mechanism is different. First let us
concentrate on the low-field regime where we define the MR as
(R$_{\uparrow\downarrow}$ - R$_{\uparrow\uparrow}$)/ R$_{\uparrow\downarrow}$ where
R$_{\uparrow\uparrow}$ and R$_{\uparrow\downarrow}$ are the resistances of the
trilayer when $\vec{M_1}$ and $\vec{M_2}$ are parallel and antiparallel
respectively. The variation of MR with R$_{\uparrow\downarrow}$ at a fixed T(15K)
with variable I, and at several Ts across the transition at constant I is shown in
Fig. \ref{mrap}. A remarkable universality of the dependence of MR emerges on the
ground state resistance of the structure. The magnetoresistance starts with a
negligibly small value at T $>$ T$_c$ but then diverges on entering the
superconducting state. While an enhancement in MR has been seen with cleaner
spacers\cite{book}, the regime of diverging MR is only accessible with a
superconducting spacer. Unlike the case of free-electron metal spacers, where the
strength of MR is attenuated by spin flip scattering in the interior of the spacer
and at spacer - ferromagnet interfaces\cite{book}, the physics of transport of spin
polarized carriers in FM-SC-FM structures is much more challenging. Here we identify
various factors which can contribute to MR and then single out the ones, which,
perhaps are truly responsible for the behavior seen in Fig. \ref{mrap}. In the inset
of Fig. \ref{mrap} we sketch a typical MR vs H curve at T $<$ T$_c$ and mark on it
some critical points where the orientation of $\vec{M_1}$ and $\vec{M_2}$ and the
effective magnetic field seen by the SC layer change significantly. For the AF
configuration (point C) the dipolar field of LSMO in the spacer cancels out but for
the FM alignment  (point B) it adds up. Thus, strictly from the angle of
pairbreaking by the dipolar field, the SC layer should have a lower resistance in
the AF configuration. Moreover, a much stronger effect of the exchange field of FM
layers on superconductivity when $\vec{M_1}$ and $\vec{M_2}$ are parallel should
make the AF state less resistive as shown originally by deGennes\cite{Gennes}. Both
these effects are inconsistent with the observation of a higher resistance in the AF
state. However, before we rule out the effect of the dipolar field altogether, a
careful examination of the situation in the vicinity of H$_{ext}$ = 0 shows that at
the negative coercive field H$_c$, just before $\vec{M_1}$ and $\vec{M_2}$ become
antiparallel, the internal field in SC $B_{\uparrow\uparrow} = -\mu_oH_{ext} -
\mu_o(m_1^d +m_2^d)$, where $m_1^d$ and $m_2^d$ are the dipolar contribution to
magnetization in the superconductor. This is what preferably result in increase in
resistance from point B to C in the inset. However, just beyond H $>$ H$_c$ in the
AF state, the internal field (B$_{\uparrow\downarrow}$) is only $\mu_oH_{ext}$
(assuming $m_1^d \sim m_2^d$). While this sudden reduction in B$_{int}$ at H$_c$
could be responsible for the plateau seen in R(H) in the AF state, the higher
resistance in the AF state still remains a puzzle. Although one could attribute it
to piling up of spin polarized quasiparticles in SC spacer, such interpretation
would require a deeper understanding of c-axis transport in these structures where
the \cuo planes are parallel to the magnetic layers. The observation of this effect
in a low carrier density cuprate of the present study is much more intriguing
because its c-axis resistivity is insulator-like in the normal state\cite{Tournier}.

We now discuss the large positive MR in the FM configuration of $\vec{M_1}$ and
$\vec{M_2}$. The H-dependence of MR in this regime derives contributions from pair
breaking effects of spin polarized electrons injected from LSMO and of the net field
seen by the YPBCO layer. Moreover, a parasitic normal component of the field due to
misalignment will introduce vortices and a large dissipation due to flux flow. We
have estimated the contribution of sample tilt by measuring its resistance in two
configurations P and Q as shown in Fig. \ref{tilt}(a\&b). We assume that the sample
platform, instead of being on the x-y plane, has a small tilt $\delta$ away from the
y-axis. In P, the sample is mounted in such a manner that the stripe of film is
nominally along \^y. Fig. \ref{tilt}(b) shows the 90$^o$ geometry such that the
stripe is now along the \^x. We rotate $\vec{H}$ in the xy-plane and measure R as a
function of the angle $\theta$ between \^y and the field direction. We expect three
distinct contributions to R($\theta$) coming from; i) Vortex dissipation due to
normal component of the field [($\Delta$R)$_{\upsilon\bot}$], ii) Lorentz force on
Josephson vortices in the plane of the film ($\Delta$R)$_{\upsilon\|}$, and iii),
the AMR of LSMO layers ($\Delta$R)$_{AMR}$ which peaks when I is perpendicular to
the in-plane field\cite{Mandal}. While all these contributions to R are periodic in
$\theta$ with a periodicity of $\pi$, in configuration P
($\Delta$R)$_{\upsilon\bot}$ will peak at $\theta = 0$ and $\pi$ where as the peak
in ($\Delta$R)$_{\upsilon\|}$ and ($\Delta$R)$_{AMR}$ will appear at $\theta =
\pi/2$ and $3\pi/2$. Since the resistivity of the sample in configuration `P' peaks
at $\pi/2$ and $3\pi/2$ (see Fig. \ref{tilt}(c)), it is evident that
($\Delta$R)$_{\upsilon\bot} <$ (($\Delta$R)$_{\upsilon\|} + $ ($\Delta$R)$_{AMR}$).
For configuration Q, on the other hand, ($\Delta$R)$_{\upsilon\bot}$,
($\Delta$R)$_{\upsilon\|}$ and ($\Delta$R)$_{AMR}$ are all in phase with peak value
appearing at $\theta = 0$ and $\pi$ as seen in Fig. \ref{tilt}(c). Clearly, the
difference of the peak height at $\theta$ = 0 of Q and $\theta$ = $\pi/2$ of P gives
us the flux flow resistance due to motion of vortices nucleated by a non-zero tilt.
Its contribution to resistance is $\sim$ 10\% at 15 K and 3kOe nominally parallel
field. Of course its strength will also vary with current. It is clear that a much
larger contribution to +ve MR comes from the in-plane field and its attendant
effects.

In conclusion, we have examined the role of pair breaking interactions such as the
dipolar field, exchange field and spin polarized quasiparticles, and of vortices in
setting the large magnetoresistance of a FM-SC-FM heterostructure at T$<$T$_c$ of
the SC layer. At low fields, we see a direct correlation between the MR and the
resistance of the ground state where the FM layer are coupled antiferromagnetically.

This research has been supported by a grant from the Board of Research in Nuclear
Sciences, Govt. of India.

\begin{widetext}
\newpage
\begin{figure} \centerline{\includegraphics[width=12cm,
angle=0]{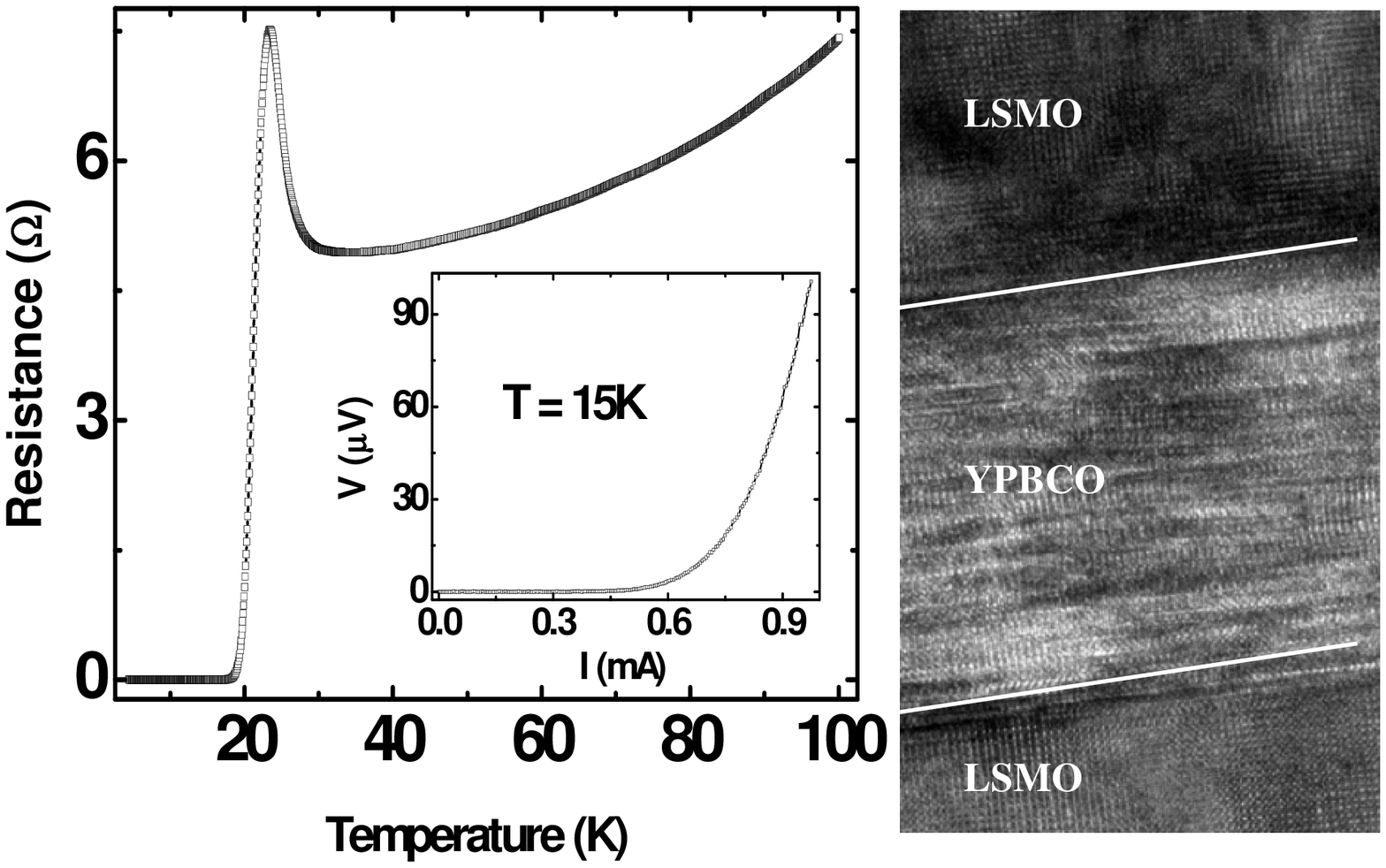}} \caption{R vs T curve for a LSMO - YPBCO - LSMO trilayer with
100nm YPBCO sandwiched between 30nm each of LSMO layers is shown in the left panel.
The onset of hole localization in YPBCO before the superconductivity sets in is
evident from the rise in resistivity near $\sim$30K. The inset of left panel shows a
typical I vs V characteristic of the trilayers at 15K. The right panel shows a high
resolution cross sectional TEM of the trilayer with d$_{YPBCO} = 200$\AA.}
\label{rtsem}
\end{figure}

\newpage
\begin{figure}
\centerline{\includegraphics[height=15cm,angle=-90]{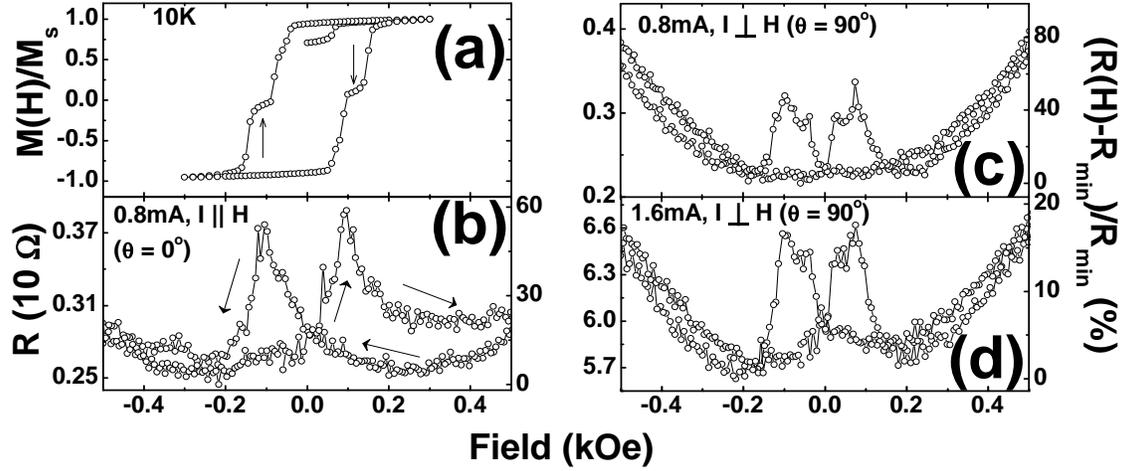}} \caption{Panel(a)
shows M vs H loop of the trilayer taken at 10K. The two symmetric small
plateaus(indicated by arrows), with zero magnetization show antiferromagnetic
coupling between the two FM layers. Panel (b) to (d) show MR measured at 15K. The
left hand side of y-axis shows resistance in units of 10$\Omega$ and the right hand
y-axis shows the $(R(H) - R_{min})/R_{min}$ in \%. In Panel (b) the current was
parallel to field where as for the remaining two Panels, the in-plane field was
orthogonal to the current ($\theta = 90^o$) which takes two values; 0.8 and 1.6 mA
for (c) and (d) respectively.} \label{rh15k}
\end{figure}

\newpage
\begin{figure} \centerline{\includegraphics[height=15cm,angle=-90]{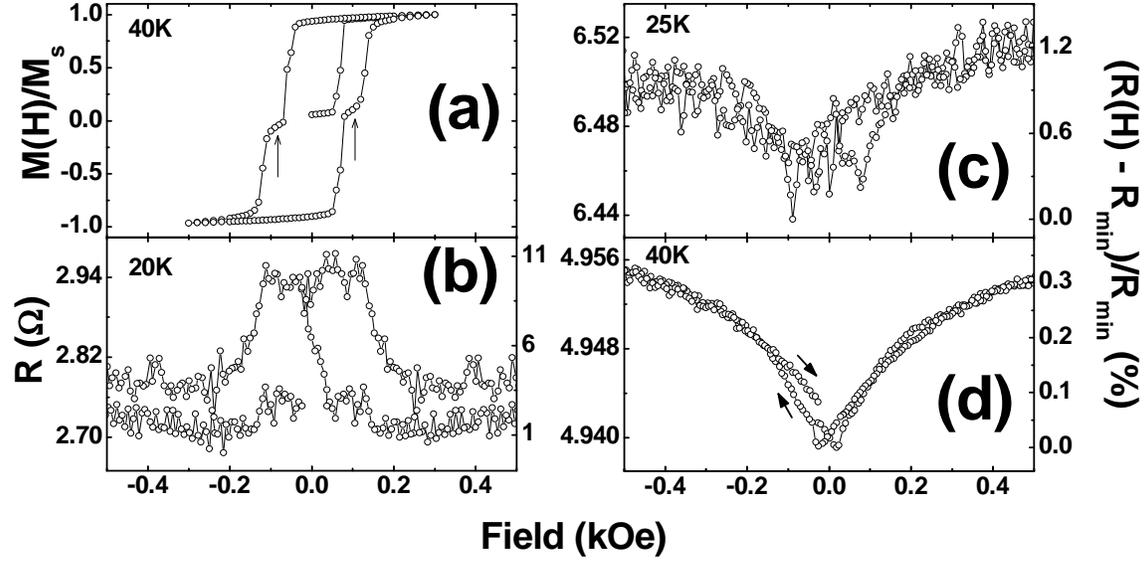}}
\caption{(a)M vs. H loop for the trilayer taken at 40K. Two symmetric small plateaus
(indicated by arrows) with zero magnetization show antiferromagnetic coupling
between the two FM layers. Panels (b), (c) \& (d) show field dependence of
magnetoresistance in $\theta = \pi/2$ ($\vec{I} \bot \vec{H}$) configuration at 20,
25 and 40K respectively. The left hand side of y-axis shows resistance ($\Omega$)
and the right hand y-axis shows the $(R(H) - R_{min})/R_{min}$ in \%.}
\label{rhvart} \end{figure}

\newpage
\begin{figure}
\centerline{\includegraphics[width=4.5in,angle=0]{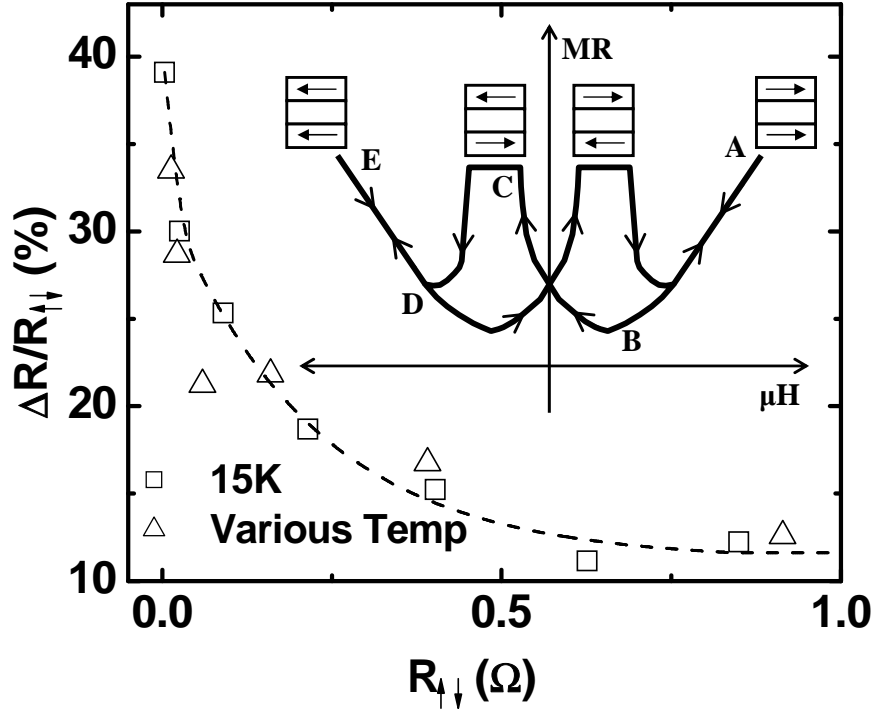}} \caption{Dependence of
MR on R$_{\uparrow\downarrow}$. This figure contains MR data collected at 15K with
variable current and at several temperatures across the transition at constant
current. A remarkable universality of the dependence of MR emerges on the ground
state resistance of the structure. The inset shows a typical sketch of MR vs H curve
in the superconducting state and identify some critical points where $\vec{M_1} \&
\vec{M_2}$ change their orientation(More details in text).} \label{mrap}
\end{figure}

\newpage
\begin{figure}
\centerline{\includegraphics[width=15cm,angle=0]{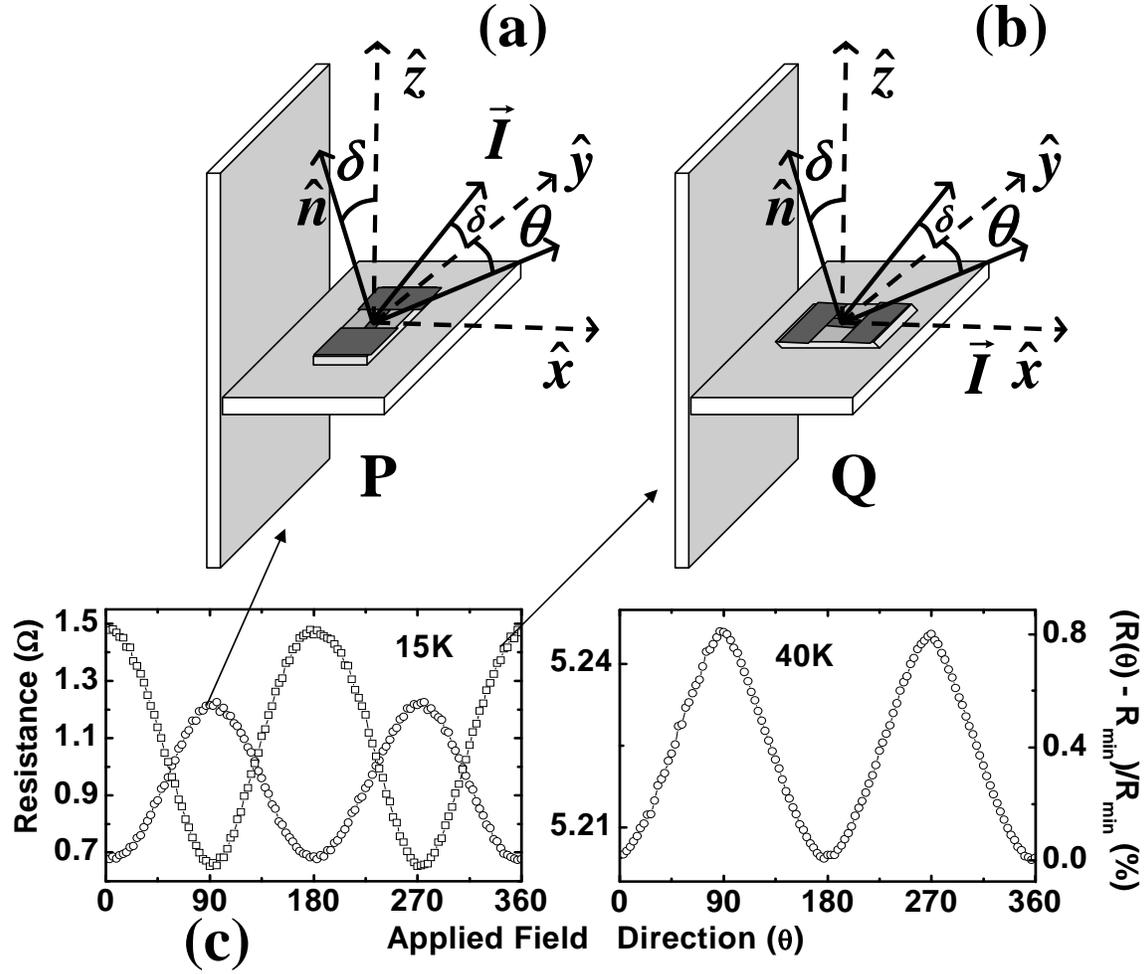}} \caption{(a) \& (b)
respectively show the configuration P and Q of sample mounting in the cryostat. The
sample stage has a non-zero tilt($\delta$) with respect to the x-y plane.(c)AMR of
the trilayer measured at 15K in configuration P and Q, and at 40K in configuration
P.} \label{tilt}
\end{figure}
\end{widetext}
\end{document}